\documentclass[%
 reprint,
showpacs,
 amsmath,amssymb,
 aps,
prl
]{revtex4-1}

\usepackage{graphicx}
\usepackage{dcolumn}
\usepackage{bm}

\begin{document}

\preprint{APS/123-QED}

\title{Bell-state correlations in current of local Fermi liquid}

\author{Rui Sakano$^{1}$}
\email{sakano@issp.u-tokyo.ac.jp}
\author{Akira Oguri$^2$}%
\author{Yunori Nishikawa$^2$}
\author{Eisuke Abe$^3$}
\affiliation{%
 $^1$Institute for solid state physics, the university of Tokyo, 5-1-5 Kashiwanoha, Kashiwa, Chiba, 277-8581 Japan \\
 $^2$Department of physics, Osaka city university, 3-3-138 Sugimoto Sumiyoshi-ku, Osaka-shi, 558-8585 Japan \\
 $^3$Spintronics Research Center, Keio University, 3-14-1 Hiyoshi, Kohoku-ku, Yokohama 223-8522, Japan
}%




\date{\today}

\begin{abstract}
We study Bell-state correlations
for quasiparticle pairs excited in nonlinear current through a double quantum dot in the Kondo regime.
Exploiting the renormalized perturbation expansion in
the residual interactions
of the local Fermi liquid
and Bell's inequality for cross correlation of spin currents through distinct conduction channels,
we derive
an asymptotically exact form of Bell's correlation
for the double dot at low bias voltages.
We find that
pairs of quasiparticles and holes excited by the residual exchange interaction
can violate Bell's inequality for the cross correlations of the spin currents.
\end{abstract}

\maketitle


Quantum dots with magnetic moments
that strongly interact with conduction electrons in connected lead electrodes
exhibit the Kondo effect,
which has been a central issue of the condensed matter physics
over the 50 years
\cite{HewsonBook}.
The low energy properties of the Kondo effect
are described well by the local Fermi liquid theory.
The local Fermi liquid is an extension of Landau's Fermi liquid
to cover quantum impurities,
in which free quasiparticles and residual interactions
account for the underlying physics
\cite{Nozieres1974,doi:10.1143/PTP.53.970,PhysRevB.64.153305,PhysRevB.92.075120,PhysRevLett.120.126802,PhysRevB.97.045406,PhysRevB.98.075404}.
In electric current through the Kondo dot
at low applied bias voltages,
residual interactions excite quasiparticle pairs
that have an effective charge of $2e$
\cite{PhysRevLett.97.016602,PhysRevLett.97.086601,PhysRevLett.100.036603,PhysRevLett.100.036604,PhysRevB.80.155322,doi:10.1143/JPSJ.79.044714,PhysRevLett.108.266401,PhysRevB.92.075120}.
This doubly-charged state
has been observed
as enhancements of the shot noise
\cite{PhysRevB.77.241303,NaturePhys5.208,PhysRevLett.106.176601,NatPhys.12.230,PhysRevB.83.075440,PhysRevB.83.241301,PhysRevLett.118.196803}.

This letter will
explore the nature
of the correlation between the quasiparticles
that are excited by the residual interactions
within the current.
In a previous work of ours
\cite{PhysRevB.97.045127},
we found that
the residual exchange interaction of a quantum dot excites
spin-entangled quasiparticles and holes
in a nonlinear current.
We investigate the nature of the spin entanglement by exploiting Bell's inequality.
Bell's theorem draws an essential distinction
between the correlations found in quantum mechanics and those found in classical mechanics.
As a {\it no-go theorem}, Bell's theorem places limits on physical possibility
\cite{PhysRevLett.28.938,PhysRevLett.49.1804,PhysRevLett.81.5039,Scheidl16112010,Nature497.7448,PhysRevLett.111.130406,Nature526.7575}.
Bell-state correlation of electrons involved in tunneling currents
through mesoscopic devices has been studied for the past 20 years
\cite{PhysRevLett.91.147901,PhysRevLett.91.157002,PhysRevLett.92.026805,PhysRevB.71.045306}.
Several studies have focused on Bell-state correlations of electrons that are entangled by many-body effects.
For example, Bell-state correlations of superconducting electron pairs have been studied 
with the Cooper pair splitter formed by a Y-shaped junction of superconductor and semiconductor
\cite{doi:10.1143/JPSJ.70.1210,PhysRevB.66.161320,Nature.461.960}.
Bell-state correlations have also been predicted
for electrons scattered
by the Kondo exchange interaction
at temperatures
near the Kondo temperature
\cite{PhysRevLett.87.277901}.
In this letter,
we propose a new way to investigate the entanglement of the Kondo state using Bell's test.

\begin{figure}[tb]
	\includegraphics[width=5.5cm]{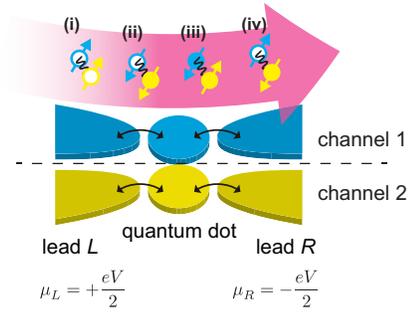}
	\caption{\label{fig:mdl}
	Schematic of the double dot and quasiparticle pairs excited within the two channels of the current.
	The filled and unfilled circles represent quasiparticles and holes, respectively, and
	the arrows attached to them indicate their spin degrees of freedom.
	Cyan and yellow indicate channels 1 and 2, respectively.
	}
\end{figure}

\paragraph{Model}
Consider the double dot illustrated in Fig. \ref{fig:mdl}.
The system is described by the action of the Anderson impurity model given as
${\cal S} =
\sum_{\mu} \int_{-{\cal T}/2}^{{\cal T}/2} dt \left( \sigma_3^{}\right )^{\mu\mu} {\cal L}_{\rm A}^{\mu}
$,
where the Lagrangian is given as
${\cal L}_{\rm A}^{\mu} =
{\cal L}_{0}^{\mu} + {\cal L}_{\rm T}^{\mu} + {\cal L}_{\rm I}^{\mu}$
with
\begin{eqnarray}
	{\cal L}_{0}^{\mu}  &=&
	\sum_{\alpha m \sigma} \int_{-D}^{D} d\varepsilon \,
	\bar{c}_{\varepsilon \alpha m \sigma}^{\mu} 
	\left( i\frac{ \partial}{\partial t}-\varepsilon \right)
	c_{ \varepsilon \alpha m \sigma}^{\mu} 
	\nonumber \\
	&& \qquad
	+ \sum_{m \sigma}
	\bar{d}_{m\sigma}^{\mu} 
	\left( i \frac{\partial}{\partial t} - \epsilon_{\rm d}^{} \right)
	d_{m\sigma}^{\mu}  \, , \\
	{\cal L}_{\rm T}^{\mu} &=&
	\sum_{\alpha m\sigma}
	\left[
	v \bar{d}_{m\sigma}^{\mu} \, \psi_{\alpha m \sigma}^{\mu}
	+ v_{}^{*} \bar{\psi}_{\alpha m \sigma}^{\mu} \, d_{m\sigma}^{\mu} 
	\right]
	\, , \\
	{\cal L}_{\rm I}^{\mu} &=&
	U \sum_{m} n_{{\rm d}m\uparrow}^{\mu} \, n_{{\rm d}m\downarrow}^{\mu}
	+ W n_{{\rm d}1}^{\mu}  \, n_{{\rm d}2}^{\mu}
	+ 2J {\bm S}_{{\rm d}1}^{\mu} \cdot {\bm S}_{{\rm d}2}^{\mu} \, .
	\label{eq:L-q-int}
\end{eqnarray}
Here, ${\cal T}$ is a long time interval,
$\sigma_3^{} =((1,0)^t, (0,-1)^t)$
is the third element of the Pauli matrix ${\bm \sigma}$,
and
the superscripts $\mu= -$ and $+$ represent the forward and backward paths of the Keldysh contour, respectively.
Note that, throughout this letter,
the time argument $t$ in the Lagrangian and the Grassmann numbers are suppressed. 
${\cal L}_0^{\mu}$ represents electrons
in the lead electrodes and the double dot.
$c_{\alpha\varepsilon m\sigma}^{\mu}$ and $\bar{c}_{\alpha\varepsilon m\sigma}^{\mu}$
are the Grassmann numbers
for electrons with spin $\sigma = \uparrow, \downarrow$ and energy $\varepsilon$
in the conduction band of the left and right leads $\alpha = L, R$ of channel $m=1,2$.
$d_{m\sigma}^{\mu}$ and $\bar{d}_{m\sigma}^{\mu}$
are the Grassmann numbers for electrons with spin $\sigma$
in level $\epsilon_{\rm d}^{}$ of dot $m$.
${\cal L}_{\rm T}^{\mu}$ represents
electron tunneling between the leads and the dots.
They are connected by tunneling matrix element $v$
through
$\psi_{\alpha m \sigma}^{\mu}
:= \int_{-D}^{D} d\varepsilon \sqrt{\rho_{\rm c}^{}} c_{\varepsilon \alpha m \sigma}$
and $\bar{\psi}_{\alpha m \sigma}^{\mu}
:= \int_{-D}^{D} d\varepsilon \sqrt{\rho_{\rm c}^{}} \bar{c}_{\varepsilon \alpha m \sigma}^{\mu}$,
where $D$ is the half width of the conduction band
 and $\rho_{\rm c}^{}=\frac{1}{2D}$ is the density of state for the conduction electrons.
Electron tunneling causes an intrinsic linewidth of the dot levels to be
 $\Gamma = 2 \pi \rho_{\rm c}^{} \left| v \right|^2$.
${\cal L}_{\rm I}^{\mu}$ represents interactions between the electrons in the double dot.
$U$ and $W$ are the intra- and interdot Coulomb interactions, respectively, and
$J$ is the exchange interaction.
The Grassmann number corresponding
to the electron occupations and the total spin in dot $m$
are given by
$n_{{\rm d}m\sigma}^{} = \bar{d}_{m\sigma}^{} d_{m\sigma}^{},
n_{{\rm d}m}^{} = \sum_{\sigma} n_{{\rm d}m\sigma}^{}$,
and
${\bm S}_{{\rm d} m}^{}
= \frac{1}{2} \sum_{\sigma \sigma'} \bar{d}_{m\sigma}^{} {\bm \sigma}_{\sigma \sigma'} d_{m\sigma'}^{}$.
We impose
the particle-hole symmetry $\epsilon_{\rm d}^{} = - \frac{U}{2} -W$
and the absolute zero temperature $T=0$
to eliminate the thermal and partition noises
and
maximize the effect of
$J$.
The bias voltage $eV$ is applied symmetrically:
 the chemical potentials of the left and right leads are
$\mu_{L}^{} = +\frac{1}{2}eV$ and
$\mu_{R}^{}=- \frac{1}{2}eV$, respectively. 
With no loss of generality,
a positive bias voltage $eV>0$ can be assumed.
We also use the natural units $\hbar=k_{\rm B}^{}=1$.

\paragraph{Bell's inequality for current correlations}

We investigate quasiparticles that become correlated across
the two channels.
In the original
argument of
Bell's theorem,
the spin correlation of two particles was studied
\cite{RevModPhys.86.419}.
However, one-by-one detection of every spin of the quasiparticles
in a quantum-scale current
is still difficult to be achieved in solid-state devices.
Thus, we exploit Bell's inequality
for two correlated currents,
derived by Chtchelkatchev {\it et al.}
\cite{PhysRevB.66.161320}.
This approach is outlined below.

The key idea of Bell's theorem is that
determinism
with a hidden variable
is assumed to
describe
any correlations
in the world.
The violation of this assumption gives a sufficient condition for the quantum entanglement.
For our double dot,
the correlation between channel 1 and 2 are assumed to be described by
a
hidden variable $\eta$.
Then, the density matrix of the whole system
can be written in the form 
\begin{eqnarray}
	\rho_{\rm HVT}^{} =
	\int d\eta \, f ( \eta ) \rho_1^{} ( \eta ) \otimes \rho_2^{} ( \eta ) \, ,
	\label{eq:DM-HVT}
\end{eqnarray}
where the distribution function for the hidden variable is satisfied with
$f( \eta ) \ge 0$ and $\int d \eta \, f( \eta) =1$,
and
$\rho_m^{} (\eta )$ is 
the density matrix for channel $m$.

Integration of the current
can give
the average of spin angled to $\theta$ per a particle in the current of channel $m$
in a long time interval from $t-\frac{\cal T}{2}$ to $t+\frac{\cal T}{2}$:
\begin{eqnarray}
	\bar{A}_{m \theta} ( t, \eta )
	=  \frac{\int_{t-\frac{\cal T}{2}}^{t + \frac{\cal T}{2}} dt' \, {\rm tr} [ \rho_m^{} ( t', \eta ) J_{m  \theta}^{\rm s} ( t' ) ]}
	{\int_{t-\frac{\cal T}{2}}^{t + \frac{\cal T}{2}} dt' \, {\rm tr} [ \rho_m^{} ( t', \eta ) J_{m}^{\rm c} ( t' ) ] }
	 \, .
\end{eqnarray}
$J_{m\theta}^{\rm s} = J_{m\theta}^{} - J_{m\theta+\pi}$ and $J_{m}^{\rm c} = J_{m\theta}^{} + J_{m\theta+\pi}$
are the spin and charge current, respectively,
where $J_{m\theta}^{}$ is the current with spin angled to the $\theta$ direction in channel $m$.
For a current which effectively carries the spin correlation,
the average spin is normalized as
$\left|  \bar{A}_{m \theta} ( t, \eta )  \right| \le 1$.
Then, the conventional derivation of Bell's inequality for two incident entangled particles
is applicable to the averaged spin in the currents through the two channels. 
We obtain the Clauser-Horne-Shimony-Holt Bell's inequality for two correlated currents as
\begin{eqnarray}
	0\le{\cal C} \le 2 \, ,
	\label{eq:CHSH-BI}
\end{eqnarray}
where the Bell's correlation is given in the form
\begin{eqnarray}
	{\cal C} =
	\left|
	F ( \theta, \varphi ) - F ( \theta', \varphi )
	+ F( \theta, \varphi' ) + F( \theta', \varphi' )
	\right|
	\, .
	\label{eq:CHSH-Bell-Correlator}
\end{eqnarray}
Here,
$F ( \theta, \varphi )=
h_{}^{\rm s}( \theta, \varphi )
/h_{}^{\rm c}$
is given by a cross-correlation of the spin current
\begin{eqnarray}
	h_{}^{\rm s} ( \theta, \varphi ) :=
	\int dt \, \left\langle  J_{1\theta}^{\rm s} (t)  J_{2\varphi}^{\rm s} (0) \right\rangle_{\rm HVT}^{}
	\, , \label{eq:scc-HVT}
\end{eqnarray}
and that of the charge current,
\begin{eqnarray}
	 h_{}^{\rm c} :=
	\int dt \, \left\langle  J_{1}^{\rm c} (t)  J_{2}^{\rm c} (0) \right\rangle_{\rm HVT}^{}
	\, , \label{eq:ccc-HVT}
\end{eqnarray}
with the average by the density matrix of the hidden variable theory,  
$\langle \cdots \rangle_{\rm HVT}^{} := {\rm tr}\left[ \rho_{\rm HVT}^{} \cdots \right]$.
Therefore,
violation of Eq. (\ref{eq:CHSH-BI}) for Bell's correlation ${\cal C}_{\rm QM}^{}$ calculated
with the fully quantum mechanical density matrix $\rho_{\rm QM}^{}$
gives a sufficient condition for quantum correlation.

{\it Current correlations---}
In our dot,
the current of the electrons with spin angled to the $\theta$ direction
in channel $m$ is given as
$I_{m\gamma} = -i ( v \bar{d}_{m\gamma}^{} \psi_{R m\gamma}^{}
	+  v_{}^{*} \bar{\psi}_{R m\gamma}^{} d_{m\gamma}^{} )$,
To calculate current correlation,
we introduce the source term
\begin{eqnarray}
	{\cal L}_{\rm sou}^{\mu} ({\bm \lambda}) =
	-i \sum_{m\gamma}
	\left( e_{}^{i \lambda_{m\gamma}^{\mu}} v \bar{d}_{m\gamma}^{} \psi_{R m\gamma}^{}
	+ e_{}^{- i\lambda_{m\gamma}^{\mu}} v_{}^{*} \bar{\psi}_{R m\gamma}^{} d_{m\gamma}^{} \right)
	.
	\nonumber \\
\end{eqnarray}
in ${\cal L}_{\rm A}^{\mu}$.
Here, $\lambda_{m\gamma}^{\mu}=(\sigma_3^{})^{\mu\mu} \lambda_{m\gamma}^{}$
is a contour-dependent source field,
and $\gamma(=\theta,\theta+\pi)$ is the spin index defined with respect to the $\theta$ direction.
The Grassmann number for an electron in the dot with spin $\gamma$
can be given
by a rotational transformation
as
$d_{m\theta}^{\mu} = \cos\frac{\theta}{2} d_{m\uparrow}^{\mu} - \sin \frac{\theta}{2} d_{m\downarrow}^{\mu}$.
$\bar{d}_{m\theta}^{\mu}, \psi_{\alpha m\theta}^{\mu}$, and $\bar{\psi}_{\alpha m\theta}^{\mu}$ are also defined
in the same manner.
Current correlations can be calculated by differentiating the generating function
$\ln {\cal Z}({\bm \lambda})$
with the corresponding source fields.
The partition function
is given in the form
\begin{eqnarray}
	{\cal Z} ( {\bm \lambda} )
	= \int 
	{\cal D}\left(\bar{c}_{\varepsilon \alpha  m\sigma}^{} \right) {\cal D} (c_{\varepsilon \alpha m\sigma}^{} )
	{\cal D} (\bar{d}_{m\sigma}^{} ) {\cal D} (d_{m\sigma}^{} )
	e^{i{\cal S} ( {{\bm \lambda} })}
	\nonumber \\
	\label{eq:prttnfnctn}
\end{eqnarray}
with
${\cal S} ( {\bm \lambda }) =
	\sum_{\mu }\int_{-{\cal T}/2}^{{\cal T}/2} dt \, (\sigma_3^{})_{}^{\mu\mu} 
	\left[
	{\cal L}_{\rm A}^{\mu}  + {\cal L}_{\rm sou}^{\mu} ( {\bm \lambda })
	\right]
$
\cite{PhysRevB.97.045127}.

\paragraph{Renormalized perturbation theory}
To take electron correlations
into account,
we use the renormalized perturbation theory
\cite{PhysRevLett.70.4007,0953-8984-13-44-314,PhysRevB.82.115123}.
At low energies
perturbation expansion
in
$U, W$, and $J$,
provides an exact result if all the terms in the series are accounted for.
However, this expansion is difficult, except for some special cases.
Below, employing the idea of the renormalized perturbation theory,
we reorganize the perturbation expansion
and effectively carry out all-order calculations at low energies.

First, we formulate the quasiparticle's Lagrangian $\widetilde{\cal L}_{\rm qp}^{\mu}$
by replacing $\epsilon_{\rm d}^{}, v, U,W,J, d_{m\sigma}^{\mu}$,
and $\bar{d}_{m\sigma}^{\mu}$ of ${\cal L}_{\rm A}^{\mu}$
with the renormalized parameters and the Grassmann numbers of the quasiparticle given by
$\tilde{\epsilon}_{\rm d}^{}, \tilde{v}, 
\widetilde{U},\widetilde{W},\widetilde{J}, \tilde{d}_{m\sigma}^{\mu}$,
and  $\tilde{\bar{d}}_{m\sigma}^{\mu}$.
These renormalized parameters and Grassmann numbers that are defined
by sets of perturbation series given
by the self-energy and the four vertex at $T=eV=0$
\cite{PhysRevB.97.045127}.
Note that the renormalized linewidth
given by $\widetilde{\Gamma}:= 2\pi \rho_{\rm c}^{}|\tilde{v}|^2$ corresponds to the characteristic energy scale,
namely, the Kondo temperature:
$T_{\rm K} = \pi \widetilde{\Gamma}/4$.
We can evaluate
$\tilde{\epsilon}_{\rm d}^{}, \widetilde{\Gamma}, \widetilde{U}, \widetilde{W}$, and $\widetilde{J}$
by using the numerical renormalization group (NRG) approach
\cite{Hewson2004,PhysRevB.82.115123,PhysRevLett.108.056402}.
The nonequilibrium effects at low bias voltages $eV \ll T_{\rm K}^{}$
arise through perturbation expansions in the residual interactions.

As a part
of the interaction effects are taken into account {\it ab initio} in the quasiparticle's Lagrangian
during renormalized perturbation expansion,
a counter term has to be introduced to avoid overcounting in the perturbation expansion.
In the other words, the total Lagrangian has to be satisfied with
${\cal L}_{\rm A}^{\mu} = \widetilde{\cal L}_{\rm qp}^{\mu}  + {\cal L}_{\rm CT}^{\mu} $.
The counter term ${\cal L}_{\rm CT}^{\mu}$,
can be expressed in terms of
the renormalized parameters and the renormalized Grassmann numbers,
which are determined
by the renormalized condition for the renormalized self-energy
 and the renormalized four-vertex.
In the particle-hole symmetric case,
the perturbation expansion up to only the second order in the residual interactions
provides an asymptotically exact form of the self-energy at $T=0$ up to the second order in $\omega$ and $V$,
and asymptotically exact forms of currents and current correlations up to order $V^3$.
Then, any higher orders terms in the residual interactions
do not yield contribution of order $V^3$
\cite{doi:10.1143/JPSJ.74.110,PhysRevB.83.075440,PhysRevB.83.241301,PhysRevLett.108.266401}.
We shall calculate the current correlations
using perturbation expansion in the residual interactions.

\paragraph{Results and discussion---}
Let us calculate ${\cal C}_{\rm QM}^{}$ in terms of the quasiparticle parameters.
Since $\left\langle I_{m\theta}^{\rm s} \right\rangle=0$ in our model,
the correlation of the spin currents given by Eq. (\ref{eq:scc-HVT})
can be rewritten into the correlation of spin current fluctuations
$\delta I_{m\theta}^{\rm s}= I_{m\theta}^{\rm s} -\left\langle I_{m\theta}^{\rm s} \right\rangle$
as
\begin{eqnarray}
	h_{\rm QM}^{\rm s} ( \theta, \varphi )
	= \lim_{{\cal T}\to \infty} \int_{-\frac{\cal T}{2}}^{\frac{\cal T}{2}} dt
	\, \left\langle \delta I_{1\theta}^{\rm s} (t) \delta I_{2\varphi}^{\rm s} (0) \right\rangle
	\, .
\end{eqnarray}
Thus, differentiation of $\ln {\cal Z}({\bm \lambda})$ with the source fields yields
\begin{eqnarray}
	h_{\rm QM}^{\rm s} ( \theta, \varphi)
	&=& - \frac{V}{2\pi} \left( \frac{V}{\widetilde{\Gamma}} \right)^2
	\left(
	\frac{1}{4} \tilde{j}^2 - \frac{1}{3} \tilde{w}  \tilde{j}
	\right) \cos ( \theta - \varphi )
	\nonumber \\
	&& + {\cal O} \left( V^5 \right)
	\, ,
\end{eqnarray}
where $\tilde{w}=\frac{\widetilde{W}}{\pi\widetilde{\Gamma}}$ and $\tilde{j}=\frac{\widetilde{J}}{\pi\widetilde{\Gamma}}$.
Note that
the spin correlation measured by $h_{\rm QM}^{\rm s} ( \theta, \varphi)$ comes from
only a portion of the entangled quasiparticle pairs within the current.
As seen in the specific form of $\ln {\cal Z}({\bm \lambda})$
\cite{PhysRevB.97.045127},
the residual
interactions can excite
four types of
the quasiparticle pairs
in the current (See Fig. \ref{fig:mdl}).
As TABLE \ref{tab:sgnofsccc} shows,
the spin and charge current correlations of these pairs have different signs from each other.
\begin{table}[b]
	\caption{\label{tab:sgnofsccc}%
	Signs of spin/charge current correlations of particle-particle(p-p), hole-hole(h-h) and particle-hole (p-h) pairs
	with parallel and antiparallel spins.
	The pairs excited in the current are shown in Fig. \ref{fig:mdl}.}
	\begin{ruledtabular}
		\begin{tabular}{lcc}
			 & \textrm{p-p or h-h pairs} &	
			\textrm{p-h pair} \\
			\colrule
			\textrm{parallel spin} & (i) $+/+$ & (ii) $-/-$ \\
			\textrm{antiparallel spin} & (iii) $-/+$ & (iv) $+/-$ \\
		\end{tabular}
	\end{ruledtabular}
\end{table}
Consequently, some of the spin and charge correlations due to these pairs
are independently canceled in the full current. 
Therefore, the correlation of the charge current
that effectively carries the spin current correlation must be calculated,
rather that of the full current $I_{m}^{\prime{\rm c}}$
given by $h_{}^{\rm fcc} = \int_{-\infty}^{\infty} dt \left\langle I_{1}^{\rm c} (t) I_{2}^{\rm c} (0) \right\rangle$ with
$I_{m}^{\rm c} = \sum_{\gamma} I_{m\gamma}^{}$.
The current correlation given by Eq. (\ref{eq:ccc-HVT}) can be written
in terms of current fluctuation of $I_{m}^{\prime{\rm c}}$ as
\begin{eqnarray}
	h_{\rm QM}^{\rm c}
	= H_{\rm QM}^{\rm c}
	+ {\cal T} \left\langle I_{1}^{\prime{\rm c}} \right\rangle \left\langle I_{2}^{\prime{\rm c}} \right\rangle
	\, ,
	\label{eq:qm-crrnt-crs-crrltn}
\end{eqnarray}
where
$H_{\rm QM}^{\rm c}
	= \int_{-\infty}^{\infty} dt
	\, \left\langle \delta I_{1}^{\prime{\rm c}} (t) \delta I_{2}^{\prime{\rm c}} (0) \right\rangle$
with $\delta I_{m}^{\prime{\rm c}} (t) = I_{m}^{\prime{\rm c}} (t) - \left\langle I_{m}^{\prime{\rm c}} \right\rangle$.
Although an explicit  expression of $I_{m}^{\prime{\rm c}}$ is not easy to derive,
the correlation can be evaluated readily using the terms of spin correlated carriers in $\ln {\cal Z}({\bm \lambda})$:
\begin{eqnarray}
	H_{\rm QM}^{\rm c}
	&=& - \frac{V}{2\pi} \bigg( \frac{V}{\widetilde{\Gamma}} \bigg)^2
	\left(
	\frac{1}{4} \tilde{j}^2 - \frac{1}{3} \tilde{w}  \tilde{j}
	\right)
	 + {\cal O} \left( V^5 \right)
	\, .
\end{eqnarray}
The leading term of the charge current is of the third order in the applied bias voltage,
$\left\langle I_m^{\rm c}\right\rangle \propto eV \left( \frac{eV}{\widetilde{\Gamma}}\right)^2$.
Thus there are two regions in
$h_{\rm QM}^{\rm c}$.
One is ${\cal T}^{-1} \gg eV \left( \frac{eV}{\widetilde{\Gamma}}\right)^2$,
 where
 $H_{\rm QM}^{\rm c} \ll {\cal T} \left\langle I_{1}^{\rm c} \right\rangle \left\langle I_{2}^{\rm c} \right\rangle$.
Then, the correlation function can be given simply as
$h_{\rm QM}^{\rm c}
\sim {\cal T} \left\langle I_{1}^{\rm c} \right\rangle \left\langle I_{2}^{\rm c} \right\rangle$.
This results in ${\cal C}_{\rm QM}^{} \sim 0$,
and ${\cal C}_{\rm QM}^{}$ never violates Bell's inequality in this region.
In the opposite region ${\cal T}^{-1} \ll eV \left( \frac{eV}{\widetilde{\Gamma}}\right)^2$,
the correlation of the fluctuations is dominant, namely, $H_{\rm QM}^{\rm c} \gg
{\cal T} \left\langle I_{1}^{\rm c} \right\rangle \left\langle I_{2}^{\rm c} \right\rangle$,
which leads to
$h_{\rm QM}^{\rm c}
\sim H_{\rm QM}^{\rm c}$.
Then, Bell's correlation is given in the form
\begin{eqnarray}
	{\cal C}_{\rm QM}^{}
	\sim K(\theta,\theta';\varphi,\varphi')
\end{eqnarray}
with
\begin{eqnarray}
	K(\theta,\theta';\varphi,\varphi')
	&=& | \cos(\theta - \varphi ) - \cos(\theta' - \varphi ) \nonumber \\
	&& \quad + \cos(\theta - \varphi' ) + \cos(\theta' - \varphi' ) | .
\end{eqnarray}
Since $K(\theta,\theta';\varphi,\varphi')$
is bounded in $\left[0,2\sqrt{2}\right]$,
it is concluded that the exchange interaction of the Fermi liquid can violate Bell's inequality.

\begin{figure}[tb]
	\includegraphics[width=5.0cm]{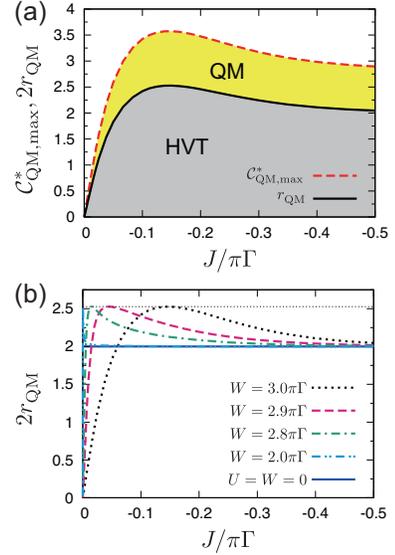}
	\caption{\label{fig:ferro}
	(Color online)
	(a) ${\cal C}_{\rm QM, max}^{*}$ and $2r_{\rm QM}^{}$
	as a function of ferromagnetic $J (<0)$
	for $U=W=3.0 \pi \Gamma$.
	The gray area is covered by the hidden variable theory,
	and the yellow area represents the sufficient condition for the quantum correlation.
	(b) $2r_{\rm QM}^{}$ as a function of ferromagnetic $J$
	for $U=3.0 \pi \Gamma$ and several choices of
	$W=3.0 \pi \Gamma,\, 2.9\pi \Gamma,\, 2.8\pi \Gamma$, and $2.0\pi \Gamma$,
	and $U=W=0$.
	The thin dotted line indicates
	the maximum value $r_{\rm QM}^{} =1+ \frac{\sqrt{21}}{3}\approx 2.528$.
	}
\end{figure}

\begin{figure}[tb]
	\includegraphics[width=5.0cm]{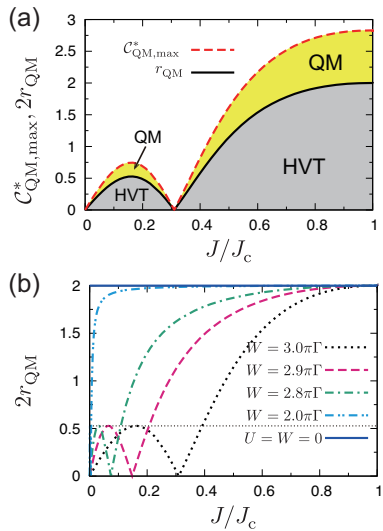}
	\caption{\label{fig:antiferro}
	(Color online)
	(a) ${\cal C}_{\rm QM, max}^{*}$ and $2r_{\rm QM}^{}$
	as a function of anitiferromagnetic $J (>0)$
	for $U=W=3.0 \pi \Gamma$.
	$J$ is normalized by the critical value $J_{\rm c}^{}$.
	The gray area is covered by the hidden variable theory,
	and the yellow area represents the sufficient condition for the quantum correlation.
	(b) $2r_{\rm QM}^{}$ as a function of ferromagnetic $J$
	for $U=3.0 \pi \Gamma$ and several choices of
	$W=3.0 \pi \Gamma,\, 2.9\pi \Gamma,\, 2.8\pi \Gamma$, and $2.0\pi \Gamma$,
	and $U=W=0$.
	The thin dotted line indicates
	the value of the local maximum $r_{\rm QM}^{} =1- \frac{\sqrt{21}}{3}\approx 0.528$.
	}
\end{figure}

However, ${\cal C}_{\rm QM}^{}$ may be difficult to measure experimentally,
because $h_{\rm QM}^{\rm c}$ is
the current correlation of the carriers that effectively carry the correlated spins.   
Next we suggest a measurable form of Bell's correlation.
It is obtained by replacing $h_{}^{\rm c}$ with the cross correlation of the full charge currents
through the two channels $h_{}^{\rm fcc}$ as
\begin{eqnarray} 
	{\cal C}_{}^* \leq 2 r
	\label{eq:rev-BI}
\end{eqnarray}
with $r = | h_{}^{\rm c}/h_{}^{\rm fcc} |$, and
\begin{eqnarray} 
	{\cal C}_{}^* 
	:= |
	F_{}^* ( \theta, \varphi ) - F_{}^* ( \theta', \varphi )
	+ F_{}^*( \theta, \varphi' ) + F_{}^*( \theta', \varphi' )
	|
	\nonumber \\
\end{eqnarray}
with
$F_{}^* (\theta,\varphi) = h_{}^{\rm s}(\theta,\varphi)/h_{}^{\rm fcc}$.
For the quantum mechanical density of states,
${\cal C}_{}^*$ and $r$ take the forms
\begin{eqnarray}
	{\cal C}_{\rm QM}^* &=& r_{\rm QM}^{} K(\theta,\theta';\varphi,\varphi')
	\, , \\
	r_{\rm QM}^{} &=& \left| \frac{h_{\rm QM}^{\rm c}}{h_{\rm QM}^{\rm fcc}} \right|
	= \left| \frac{1 - \frac{4}{3} \left( \frac{\widetilde{w}}{\widetilde{j}}\right)}
	{1 + \frac{4}{3} \left( \frac{\widetilde{w}}{\widetilde{j}}\right)^2} \right|
	\, .
\end{eqnarray}
Since ${\cal C}_{}^{*}$ is simply given by a product of ${\cal C}_{}$ and $r$,
${\cal C}_{\rm QM}^{*}$ can also violate Bell's inequality given by Eq. (\ref{eq:rev-BI}).
The maximum value of ${\cal C}_{\rm QM}^{*}$ is given by ${\cal C}_{\rm QM, max}^{*}=2\sqrt{2}r_{\rm QM}^{}$,
which corresponds to the Tselson's bound
\cite{Cirel'son1980} in our model.
This bound gives the upper limit for the correlation in the quantum regime.
${\cal C}_{\rm QM, max}^{*}$ and $2r_{\rm QM}^{}$ are plotted as a function of $J$;
$J\leq0$ and $J\geq0$ for $U=W=3.0\pi \Gamma$ in Fig. \ref{fig:ferro} (a) and Fig. \ref{fig:antiferro} (a), respectively.
A critical point appears at $J=J_{\rm c}^{}>0$
\cite{PhysRevLett.108.056402,PhysRevB.86.125134}.
For $J> J_{\rm c}^{}$,
two electrons occupying in the double dot form
an isolated singlet state and decouple
from the conduction electrons,
and  then no charge currents  can flow through the double dot.
Thus,
we focus on the region $J<J_{\rm c}^{}$,
in which the low-energy state is accounted for
by the local Fermi-liquid, and electric current flows through the dot.
The region between ${\cal C}_{\rm QM, max}^{*}$ and $2r_{\rm QM}^{}$ represents a sufficient condition
that the correlation of spin currents across the two channels is quantum mechanical in nature.

For the finite strength of $J>0$, the value ${\cal C}_{\rm QM, max}^{*}$ takes a local minimum to zero,
where the excited quasiparticle pairs with parallel and antiparallel contributions
to the spin correlation cancel each other out.   
Thus, Bell's test is not applicable with this value of $J$.
Experimentally,
the violation of Bell's inequality can be confirmed
through observation with values of ${\cal C}_{\rm QM}^{*}$ larger
than the theoretically calculated value of  $2r_{\rm QM}^{}$.
This parameter $2r_{\rm QM}^{}$ depends on the strength of $U, W,$ and $J$,
which can be evaluated using NRG calculations.
$2r_{\rm QM}^{}$ is plotted as a function of $J$ for several choices of $U$ and $W$
for $J\leq0$ and $J\geq0$ in Fig. \ref{fig:ferro} (b) and Fig. \ref{fig:antiferro} (b), respectively.

For $|J| \gg T_{\rm K}^{}$, the values of $h_{\rm QM}^{\rm fcc}$ coincide with $h_{\rm QM}^{\rm c}$,
which results in $r_{\rm QM}^{} \to 1$ and the $r_{\rm QM}^{}$ independent form of Bell's inequality is recovered.
In this region, therefore,
Bell's test can be examined without the need for any numerical calculations of $r_{\rm QM}^{}$.

Finally,
we discuss the causal locality of Bell's theorem in our model.
Bell-state correlations in our model are induced by entangled quasiparticles that are
excited by the residual exchange interaction that is scaled by $T_{\rm K}^{}$.
Therefore, for the causal locality to hold,
the two measurements in channel 1 and 2 must be separated by a distance
$d \gg c t_{\rm K}^{}$,
where $t_{\rm K}^{} = \frac{\hbar}{k_{\rm B}^{}T_{K}}$ is the Kondo time scale
and $c$ is the speed of light.
For a typical Kondo temperature of quantum dots $T_{\rm K}\sim1$K,
$d$ must be much larger than $ct_{\rm K}^{}\sim4.58\times10^{-2}$m.

RS thanks Shiro Kawabata, Taro Wakamura,
Thierry Martin, and Kensuke Kobayashi
for the helpful discussions,
and thanks Yuya Shimazaki for the inspiring discussions.
This work was partially supported by JSPS KAKENHI Grant
Numbers
JP26220711, JP15K05181, JP16K17723, and JP18K03495.


%

\end{document}